\documentclass[runningheads]{llncs}
\usepackage[T1]{fontenc}
\usepackage{graphicx}
\begin{document}
\title{Changing Pedagogical Paradigms: Integrating Generative AI in Mathematics to Enhance Digital Literacy through `Mathematical Battles with AI'}
\titlerunning{Mathematical Battles with AI}
%
\author{Moskalenko~M.~A. 
Trifanov~A.~I. 
Popkov~R.~A. 
\and
Tabieva~A.~V. 
\and
Smirnova~M.~D. 
\and
Pravdin K.~V. 
\and
Bakalin D.~A. 
}
\authorrunning{Moskalenko~M.~A. et al.}
%
\institute{ITMO University, Saint Petersburg, Russia\\  
\email{moskalenko.mary@gmail.com}}

\maketitle              
\begin{abstract}
This paper introduces `Math Battles with AI', an innovative competitive format 
designed at ITMO University to redefine the role of generative AI in mathematics 
education. Moving away from a purely defensive stance, the authors propose 
an AI agent with intentionally increased hallucination likelihood in specific 
modes to train verification skills. We describe the three-stage tournament structure 
and a specialized assessment system that rewards critical verification over blind reliance. 
Initial results indicate a significant shift in student mindsets, fostering 
essential skills in digital hygiene and prompt engineering. This work serves as a 
practical guide for academic institutions aiming to leverage AI for enhancing, 
rather than undermining, intellectual development.

\end{abstract}
\section{Introduction: educational paradigm evolution -- from prohibitions to conscious integration}

Together, the rapid technological development and ready availability of generative artificial intelligence have plunged the entire academic field into a state that can only be described as a `pedagogical challenge'. The absolute majority of universities worldwide is deeply attached to an educational system which is purely based 
on the individual assessment of both students' natural talent for the subjects and their ability to reproduce the knowledge acquired from their teachers through solving problems, proving theorems, etc. Nowadays, this time-tested, `traditional' approach is colliding with a reality where even the most difficult of questions can be answered by AI in a matter of seconds. The obvious -- and by `obvious' we mean `historically-built' -- reaction of prohibiting the use of AI resources, blocking access, and tightening controls over the originality of assignments submitted is not only ineffective but counterproductive from a long-term perspective. It creates friction between teachers and students, encouraging the invention of increasingly sophisticated ways to outsmart the system. Moreover, it ignores a simple truth: AI is set to become as essential a tool for modern specialists as computers, computer algebra systems, code editors, and calculators once were. The aforementioned  tools, especially computer algebra systems~\cite{ref_modern,ref_syktyvkar,ref_modern2}, illustrate how technologies once met with resistance were ultimately woven into the fabric of modern education. Banning AI is counterproductive, as it fails to prepare students for the future they will live and work in. Hardly anyone today thinks that university graduates should waste their time solving differential equations or writing simple scripts by hand.

A team of mathematicians from ITMO University's `TInT' megafaculty is certain that we should make a conscious decision to reshape our pedagogical approach. Rather than fight with AI, we should move toward its careful integration. Our central hypothesis suggests that one of the the goals of modern education is to teach students to use AI effectively and be productive without abandoning critical thinking. `Mathematical battles with AI' format, developed by us over the past year, embodies this philosophy. It demonstrates that AI can be a vital tool for students despite its inherent unreliability.
The aim of this pedagogical experiment is to turn the potential threat to education as we know it into a powerful resource essential for developing new skills required in the 21st century. Similar approaches, which prioritize integrating technologies instead of resisting them, can be found in the international educational agenda where the point is shifting toward learning to work with AI \cite{ref_unesco,ref_chatbot}.\\

\section{Materials and methods} 

\emph{The core principle: AI as a `difficult partner' in learning}
 \begin{enumerate}
     \item \emph{The Principle of Mindful Interaction.} We operate on the premise that students must understand the fundamental nature of AI from the very beginning: it is not an omniscient oracle, but a complex statistical tool lacking human-like comprehension. It is capable of both brilliant insights and systemic, often subtle, errors known as `hallucinations'. Our goal is to cultivate informed skepticism rather than blind trust, fostering a resilient habit of verifying any AI-generated output.
     \item \emph{The Principle of Developing Critical Thinking Through Practice.} We believe that the most valuable skill in the AI era is not the ability to find an answer quickly, but the capacity to evaluate its accuracy, logical soundness, internal consistency, and relevance to the task at hand. This skill cannot be cultivated through lectures alone; it is honed only through practice -- in situations where the cost of an error is tangible. 
     \item \emph{The Principle of Cultivating Prompt Engineering Competencies.} The quality and utility of an AI's response depend directly and decisively on the precision of the input. We aim to teach students how to engage in a structured, purposeful dialogue with AI: formulating tasks accurately, consistently refining conditions, and iteratively improving their prompts based on the system's previous outputs. This skill is widely recognized as a key competency in current research on human-AI interaction \cite{ref_arxiv,ref_wong}.
     \item \emph{The Principle of Controlled Complexity.} To achieve these goals, we have designed a specialized educational environment where AI behavior is not random, but governed by a specific scenario. The AI acts as an unreliable assistant whose strengths and weaknesses are known to the organizers but initially hidden from the participants. This approach transforms the interaction with technology into a manageable process, providing a basis for subsequent in-depth analysis and objective evaluation. Similar concepts of managed technology-enhanced learning are discussed in the context of instructional design \cite{ref_mayer}.
 \end{enumerate}

 \emph{Tournament Architecture: Phased Immersion in AI Cooperation}\\

The tournament consists of three consecutive rounds, each addressing specific pedagogical objectives and increasing the complexity of the AI interaction model. This structure ensures a gradual and systematic development of the target competencies.\\

\emph{Round 1: `The Passcode' -- Reinforcing the Foundations of Classical Education}\\ \\
The first round is intentionally AI-free, requiring teams to solve a series of simple problems. Each correct solution earns them a puzzle piece; completing the puzzle grants entry to the next stage. `The Passcode' mirrors traditional math competitions to remind participants that fundamental knowledge and core skills remain the essential foundation of any expertise \cite{ref_edu}. A team lacking a solid mathematical foundation will find itself unable not only to solve problems independently but, more crucially, to distinguish a correct AI hint from a hallucination in subsequent rounds. This stage serves as a powerful incentive for students to strengthen their own knowledge. Furthermore, by incorporating puzzle-solving and limited entry attempts, the game forces students to master strategic planning and resource management while working under pressure and high-stakes conditions.\\

\emph{Round 2: `AI Duel' -- a laboratory for critical analysis and digital hygiene}\\ \\
This pivotal round marks the direct integration of AI into the competition, bringing our pedagogical philosophy to life. Teams are not merely permitted but required to make at least fifteen queries to the AI. However, unlike a `naive' usage scenario, the scoring system is designed to penalize blind trust and mindless AI reliance while rewarding critical verification and analytical effort. This `carrot and stick' approach is managed by the AI system itself, pre-configured to evaluate the quality and effectiveness of the prompts.

\emph{Dichotomy of AI Behavior Models:} The AI operates in two distinct modes, teaching students to differentiate their level of trust in the tool based on the context and the nature of the query.

\emph{The `Advisor' Mode:} In this mode, the AI provides truthful and accurate responses regarding fundamental mathematical facts, theorems, definitions, and formulas. To some extent, the AI functions as a vast mathematical reference tool, mirroring the approach of certain competitions that permit the use of reference materials. However, when queried for a strategy or a specific solution plan, it is programmed to provide intentionally flawed but plausible logical chains. This simulates the real-world behavior of bots, where an AI, acting with high confidence, hallucinates incorrect information.

\emph{The `Calculator' Mode:} In this mode, the AI is accurate and reliable for simple arithmetic and algebraic operations. However, when performing complex, multi-step calculations -- such as limits or integrals -- it may experience `glitches' and produce incorrect results. In this capacity, the AI resembles early computer algebra systems, whose errors were only eliminated over time by developers. This teaches students not to delegate calculations to the AI mindlessly, but to maintain critical oversight of the computational process. To some extent, the participants take on the vital role of software testers.

\emph{A Differentiated Scoring System as a Tool for Pedagogical Impact:} This system lies at the heart of the format; it does more than just record results -- it actively shapes the culture of human-AI interaction:\\
\begin{itemize}
    \item Correct Solution: +5 points. The base reward for achieving the final goal.
    \item Partial Solution: +2 points. An incentive for making progress, even if the final answer is not reached.
    \item Detecting AI Deception: +2 points. A direct and powerful incentive for constant skepticism and verification. Students receive a significant reward not for the solution itself, but for their vigilance and critical thinking.
    \item Using a False AI Hint: $-1$ point. A crucial penalty for the uncritical acceptance of information and mental lethargy. The negative score serves as a clear marker of error, encouraging self-reflection.
    \item Correct Use of an AI Hint: +0.5 points. This rewards a more nuanced skill: the ability to filter out falsehoods and extract a kernel of truth even from an imperfect or misleading AI response.
    \item Penalty for Failing to Meet the AI Query Quota. This rule, paradoxical at first glance, emphasizes our core objective: not to avoid AI, but to actively learn how to collaborate with it. We compel students to step outside their comfort zone and practice interacting with a `difficult' partner. We believe this experience will be equally valuable in human-to-human collaboration.
\end{itemize}

\emph{Round 3: `Prompt Battle' -- Strategic AI Engagement Under Resource Scarcity and Open Confrontation}\\ \\

The final round closely simulates the conditions of real-world R\&D or a high-stakes innovative environment, where analytical resources are scarce and must be used with maximum efficiency. A key pedagogical breakthrough of this format is the `Reconnaissance' stage.

For 15 minutes, teams have the opportunity to ask the AI clarifying questions regarding complex problems. Crucially, all team queries (prompts) are visible to opponents in real-time via a live feed; however, the AI's responses remain confidential to the team that asked the question. This seemingly simple mechanic generates an incredibly rich palette of pedagogical and tactical scenarios:
\begin{itemize}
    \item Strategic Thinking Development: Students must consider not only what to ask the AI but also what information about their problem-solving strategy they are revealing to the opposing team. This teaches them to create prompts that maximize utility while revealing as little of their `hand' as possible.
    \item Stimulation of Analytical Activity: Teams must constantly monitor their opponents' query feed, attempting to reverse-engineer their thought processes and identify strengths or weaknesses in their understanding of the problem. This is an active, rather than passive, process.
    \item Refinement of AI Communication Skills: Under strict time constraints, students learn to write prompts with extreme precision and brevity. They recognize that every failed attempt is not just a loss of time but a tactical advantage handed to the enemy.
    \item Final Presentation and Evaluation: In the subsequent presentation of solutions, 30\% of the total score is dedicated specifically to the quality of AI interaction (sophisticated prompting, answer analysis, and error detection). This solidifies the status of AI collaboration as a distinct, high-value competency.
\end{itemize}

\section{Pedagogical Outcomes and Long-Term Impact}

The pilot testing of the `Math Battles with AI' format has revealed a range of qualitative changes in students' attitudes and approaches to problem-solving:
\begin{enumerate}
    \item  Deep Transformation of AI Perception: A shift occurs from viewing AI as a `magic black box' that generates ultimate truths to a pragmatic understanding of it as a powerful but imperfect tool. Students recognize its clearly defined strengths and weaknesses, requiring skilled handling.
    \item Sustainable `Digital Immunity': Participants develop an automatic, effortless habit of verifying key assertions, intermediate calculations, and AI-generated logical conclusions. This critical evaluation skill becomes transferable, applied to any digital source, aligning with global media literacy goals \cite{ref_hobbs}.
    \item Qualitative Growth in AI Communication Skills: A sharp evolution in prompt quality is observed from the first to the final round. Teams move from vague, general questions to specific, structured, and deeply considered queries that demonstrate a grasp of both the subject matter and the system's logic.
    \item Significant Increase in Academic Motivation: Students perceive the format as highly relevant, practice-oriented, and authentic. It directly addresses the challenges they face in their daily studies, providing practical tools for working with new technologies. This considerably boosts engagement and interest in the subject, a key factor in successful learning \cite{ref_ryan}.
\end{enumerate}

\section{Conclusion: `Math Battles with AI' -- A Blueprint for an Educational Ecosystem in the Era of Synthetic Intelligence}

`Mathematical Battles with AI' is not merely another competition; it is a live laboratory for a new educational paradigm -- one characterized not by opposition, but by a productive synthesis of human and artificial intelligence. Rather than pitting one against the other, we intentionally create pedagogical conditions for their cooperation, where human critical, creative, and strategic thinking directs, monitors, and amplifies the computational power and informational capacity of the algorithm.

This format should be viewed not as a final solution but as a successful pilot prototype that has proven its viability and effectiveness. It demonstrates that the academic and pedagogical community can and must move beyond a defensive stance toward technology. Instead, we should shape the future of education by creating robust models that prepare students for the challenges they are yet to face. The intentional and pedagogically sound integration of AI -- not just as a tool, but as an object of study and a `difficult partner' in problem-solving -- paves the way for a new, universal literacy. In this AI-driven era, the ability to engage in correct, critical, and productive interaction with artificial intelligence becomes a core competency for any modern graduate.

%
%
%

\end{document}